\documentclass[
    ,final            
  ]
  {aipproc}

\layoutstyle{6x9}

\newcommand*{\be}{\begin{equation}}
\newcommand*{\ee}{\end{equation}}

\newcommand*{\bea}{\begin{eqnarray}}
\newcommand*{\eea}{\end{eqnarray}}

\newcommand*{\pd}{\partial}
\newcommand*{\pdm}{\pd_{\mu}}
\newcommand*{\pdn}{\pd_{\nu}}

\begin{document}

\title{Finite-Temperature Yang-Mills Theory\\ in Landau Gauge}

\classification{11-10.Wx 12.38.-t 12.38.Lg 12.38.Mh 14.70.Dj}
\keywords      {Finite-temperature field theory; Quantum chromodynamics; Other nonperturbative calculations; Quark-gluon plasma; Gluons}

\author{A. Maas}{
  address={Institute for Nuclear Physics, Darmstadt University of Technology, Schlo{\ss}gartenstra{\ss}e 9, D-64289 Darmstadt, Germany}
}

\author{J. Wambach}{
  address={Institute for Nuclear Physics, Darmstadt University of Technology, Schlo{\ss}gartenstra{\ss}e 9, D-64289 Darmstadt, Germany}
  ,altaddress={Gesellschaft f\"ur Schwerionenforschung mbH, Planckstr. 1, D-64291 Darmstadt, Germany}
}

\author{B. Gr\"uter}{
  address={Institute of Theoretical Physics, T\"ubingen University, Auf der Morgenstelle 14,\\ D-72076 T\"ubingen, Germany}
}

\author{R. Alkofer}{
  address={Institute of Theoretical Physics, T\"ubingen University, Auf der Morgenstelle 14,\\ D-72076 T\"ubingen, Germany}
}

\begin{abstract}
The gluon and ghost propagators in Landau Gauge Yang-Mills Theory are investigated. Self-consistent solutions are obtained from their equations of motion above and below the presumed phase transition. Gluon confinement is manifest in these solutions and can be read off the infrared behavior of the gluon and ghost propagators.
Confinement prevails below the presumed phase transition. Above and in the infinite temperature limit, a qualitative change is observed: The chromoelectric sector exhibits screening, while long-range chromomagnetic interactions, mediated by soft modes, are still observed. At least part of the gluon spectrum is still confined. These findings agree with corresponding lattice results. 
\end{abstract}

\maketitle


It is by now well established that QCD undergoes a phase transition, or at least a rapid crossover, at some critical temperature. There is significant evidence that the equilibrium system in the high-temperature phase is a strongly interacting one. This was already anticipated by the infinite-temperature limit of the equilibrium state which is described by a confining three-dimensional (3d) theory. This has been confirmed by lattice calculations, see e.g. \cite{Cucchieri:2001tw}, and also by the present work. The latter investigates the equilibrium properties of gluons at different temperatures using Dyson-Schwinger equations (DSEs) \cite{Alkofer:2000wg}.

The following investigations are restricted to pure Yang-Mills theory, as substantial evidence exists that the non-perturbative features of QCD are generated in the gauge sector. The equilibrium theory is governed by the Euclidean Lagrangian
\bea
\mathcal{L}&=&\frac{1}{4}F_{\mu\nu}^aF_{\mu\nu}^a+\bar c^a \pdm D_\mu^{ab} c^b\nonumber\\
F^a_{\mu\nu}&=&\pdm A_\nu^a-\pdn A_\mu^a-gf^{abc}A_\mu^bA_\nu^c\nonumber\quad\quad\quad D_\mu^{ab}=\delta^{ab}\pdm+gf^{abc}A_\mu^c\nonumber~,
\eea
with the field strength tensors $F_{\mu\nu}^a$ and the covariant derivative $D_\mu^{ab}$. $A_\mu^a$ denotes the gluon field and $\bar c^a$ and $c^a$ are the ghost fields describing part of the quantum fluctuations of the gluon field. The Landau gauge is used as it is best suited for the purpose at hand \cite{Alkofer:2000wg}.

Important properties of the gluon and ghost fields are characterized by their respective two-point Green's functions, i.e.\ the propagators. Their infrared properties are linked to the presence of confinement. Especially, a particle is absent from the physical spectrum if its propagator $D(q^2)$ vanishes in the infrared \cite{Alkofer:2000wg} 
\be
\lim_{q\to 0} D(q^2)=0.\label{confcrit}
\ee
The propagators can be used to define dressing functions as
\bea
D_G(q)&=&-\frac{G(q_0^2,q_3^2)}{q^2} \; ,\nonumber\\
D_{\mu\nu}(q)&=&P_{\mu\nu}^T(q)\frac{Z(q_0^2,q_3^2)}{q^2}+P_{\mu\nu}^L(q)\frac{H(q_0^2,q_3^2)}{q^2}\nonumber~.
\eea
Here $G$ denotes the ghost dressing function, and $Z$ and $H$ the ones for gluons being transverse or longitudinal w.r.t.\ the heat bath. To obtain these functions, a truncated set of DSEs for the propagators has been solved \cite{Fischer:2002hn,Gruter:2004bb,Maas:phd,Maas:2004se}. The necessity to truncate the infinite set of coupled Dyson-Schwinger equations generates several problems concerning gauge invariance. These have been dealt with and the errors are, at least qualitatively, under control \cite{Fischer:2002hn,Maas:phd,Maas:2004se}. Recently, some of the assumptions have been confirmed by lattice calculations \cite{Cucchieri:2004sq} and semi-perturbative methods \cite{Schleifenbaum:2004id}.

Calculations have been performed at $T=0$ \cite{Fischer:2002hn}, $T<T_c$ \cite{Gruter:2004bb}, $T>T_c$ \cite{Maas:phd}, and $T\to\infty$ \cite{Maas:2004se}. Those at $T<T_c$ have used a toroidal  discretized space-time while the others were done in the continuum. Note that, for $T\to\infty$, the theory becomes dimensionally reduced to a 3d Yang-Mills theory plus an adjoint Higgs field, where $Z$ corresponds to the (chromomagnetic) 3d gluon and $H$ to the (chromoelectric) Higgs.

The ghost dressing function is found to diverge at zero momentum. This behavior is not affected qualitatively by temperature \cite{Gruter:2004bb,Maas:phd,Maas:2004se}. The divergence indicates the mediation of long-range forces which are still present even at $T\to\infty$. In addition, this divergence is connected to the confinement mechanism \cite{Alkofer:2000wg}. This is an indication for some residual gluon confinement in the high-temperature phase.

\begin{figure}[t]
\includegraphics[width=0.5\textwidth]{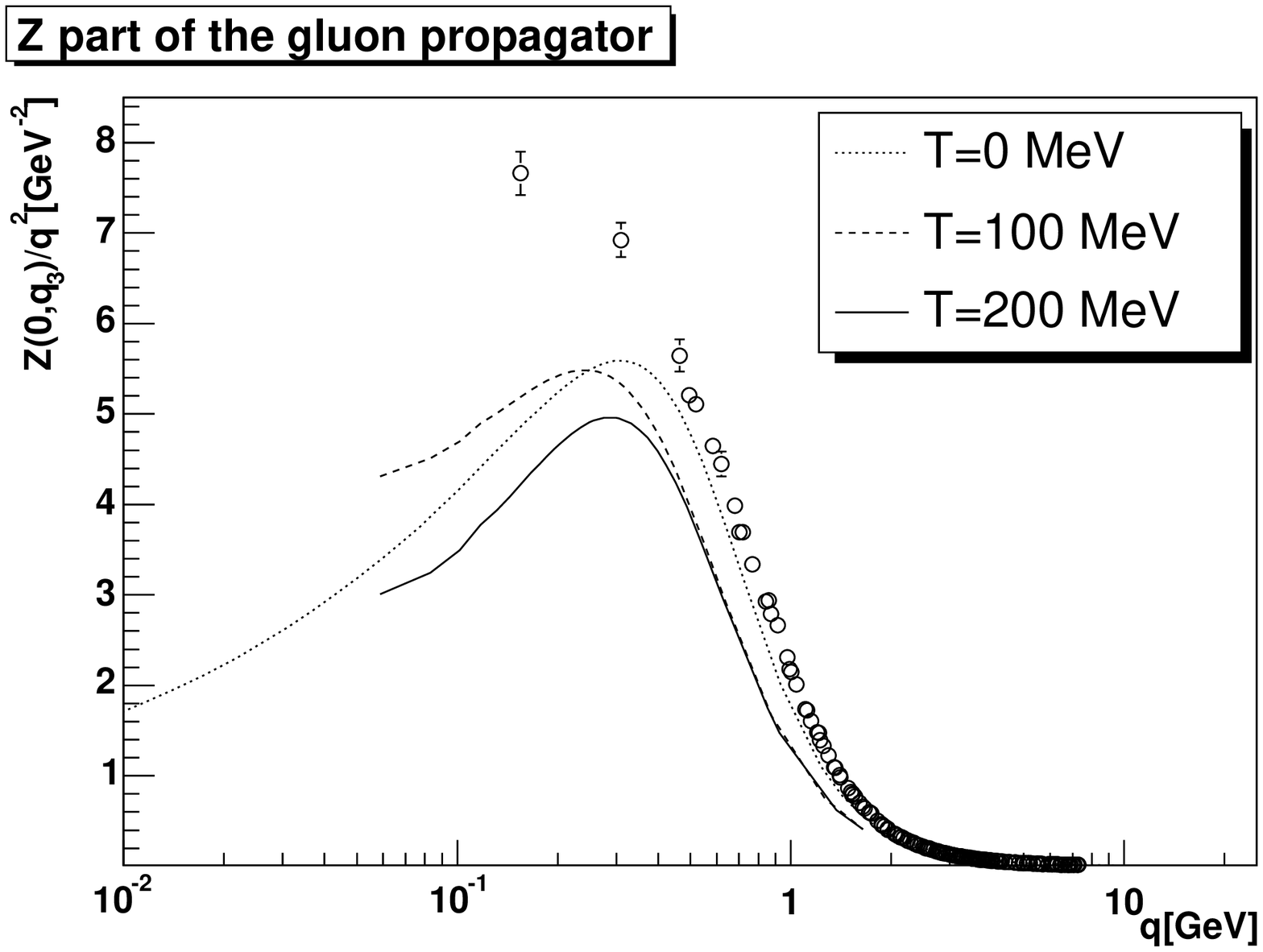}\includegraphics[width=0.5\textwidth]{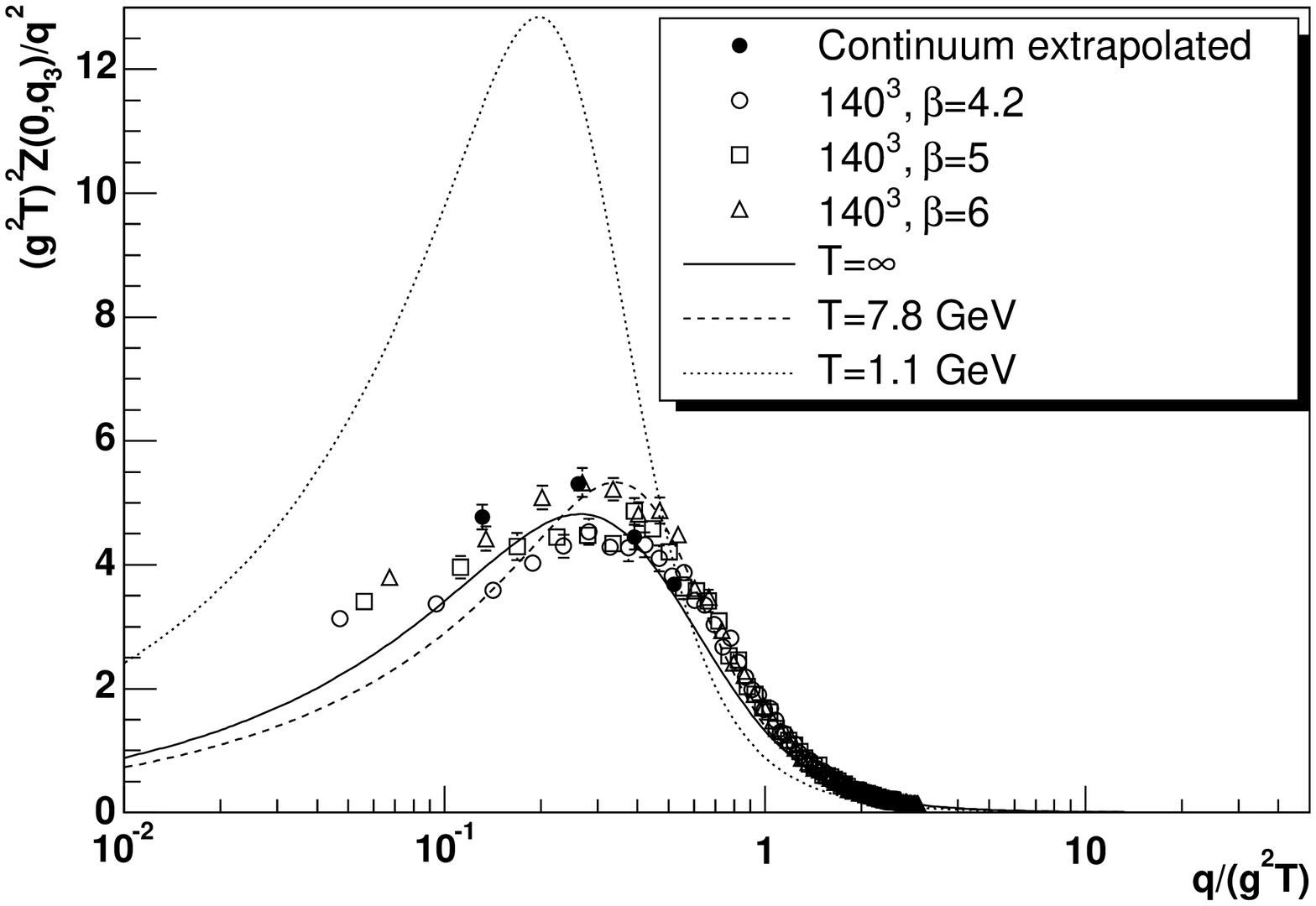}
\caption{The transverse gluon propagator: The left panel displays $T<T_c$ results compared to $T=0$ lattice data \cite{Bowman:2004jm} while the right panel shows the $T>T_c$ case compared to corresponding $T\to\infty$ lattice data \cite{Cucchieri:2001tw,Cucchieri:2003di}.
\label{gluonfig}}
\end{figure}

The results for the transverse gluon are shown in Fig.~\ref{gluonfig}. Below the phase transition, the propagator exhibits confinement by virtue of condition \eqref{confcrit}. It becomes steeper in the infrared with increasing temperature. In the high-temperature phase there is no qualitative difference. Hence, the transverse gluon is always confined. This can be interpreted as over-screening. In fact, the temperature dependence is surprisingly small. The lattice data at $T=0$ show strong finite-volume effects in the infrared, but have very recently been shown to bend over when large volumes are employed, see e.g.\ \cite{Oliveira:2004gy}, as is the case in the 3d theory, where significantly larger lattices can be used.

At $T=0$, $H=Z$ and thus gluons longitudinal w.r.t.\ the heat bath are confined as well. At $T<T_c$, $H$ becomes shallower with temperature in the infrared \cite{Gruter:2004bb}. It, however, still exhibits confinement. In the high-temperature phase, the situation is much different \cite{Maas:phd,Maas:2004se}. In contrast to the transverse sector, it shows screening and behaves similar to a massive particle. Nevertheless, a comparison to lattice results and perturbation theory indicates that the Higgs propagator contains sizeable non-trivial effects \cite{Maas:2004se}.

In conclusion, the infrared behavior of gluon and ghost propagators, employing DSEs, has been analyzed. At vanishing and small temperatures, the results show manifest gluon confinement. It is found that, even in the $T\to\infty$ limit, strong long-range correlations are present, leading to a non-perturbative behavior of the soft modes. Part of the gluons are still confined: The Gribov-Zwanziger scenario (see e.g.\ \cite{Zwanziger:2003cf}) applies at {\it all} temperatures. These results, together with the ones of lattice calculations, demonstrate that the high-temperature phase is far from trivial. Furthermore, this may have significant impact on the thermodynamic properties near the phase transition \cite{Zwanziger:2004np} and thus on experimentally accessible observables.

\begin{theacknowledgments}
A.M. thanks the organizers of Quark Confinement and the Hadron Spectrum VI for the very inspiring conference and the opportunity to present this work. This work is
supported by the BMBF under grant number 06DA116, by the European Graduate School Basel-T\"ubingen (DFG contract GRK683) and by the Helmholtz association (Virtual Theory Institute VH-VI-041).
\end{theacknowledgments}

\bibliographystyle{aipproc}   

\begin{thebibliography}{9}

\bibitem{Cucchieri:2001tw}
A. Cucchieri, F. Karsch and P. Petreczky, {\it Phys. Rev.} 
{\bf D64}, 036001 (2001).

\bibitem{Alkofer:2000wg}
R. Alkofer and L. von Smekal, {\it Phys. Rept.} 
{\bf 353}, 281 (2001).

\bibitem{Fischer:2002hn}
C. S. Fischer and R. Alkofer, {\it Phys. Lett.}
{\bf B536}, 177 (2002);
{\it Phys. Rev.}  {\bf  D67}, 094020 (2003).

\bibitem{Gruter:2004bb}
B.~Gr\"uter, R.~Alkofer, A.~Maas and J.~Wambach,
arXiv:hep-ph/0408282.

\bibitem{Maas:phd}
A.~Maas, J.~Wambach, and R.~Alkofer, in preparation.
A.~Maas, PhD thesis, TU Darmstadt, 2004.

\bibitem{Maas:2004se}
A.~Maas, J.~Wambach, B.~Gr\"uter and R.~Alkofer,
Eur.\ Phys.\ J.\ C {\bf 37} (2004) 335
[arXiv:hep-ph/0408074].

\bibitem{Cucchieri:2004sq}
A.~Cucchieri, T.~Mendes and A.~Mihara,
arXiv:hep-lat/0408034.

\bibitem{Schleifenbaum:2004id}
W.~Schleifenbaum, A.~Maas, J.~Wambach and R.~Alkofer,
arXiv:hep-ph/0411052.

\bibitem{Bowman:2004jm}
P. O. Bowman, U. M. Heller, D. B. Leinweber, M. B. Parappilly and A. G. Williams,
arXiv:hep-lat/0402032.

\bibitem{Cucchieri:2003di}
A. Cucchieri, T. Mendes and A. R. Taurines, {\it Phys. Rev.}
{\bf D67}, 091502 (2003).

\bibitem{Oliveira:2004gy}
O.~Oliveira and P.~J.~Silva,
arXiv:hep-lat/0410048 and references therein.

\bibitem{Zwanziger:2003cf}
D.~Zwanziger,
{\it Phys. Rev.} {\bf D69}, 016002 (2004)
and references therein.

\bibitem{Zwanziger:2004np}
D.~Zwanziger,
arXiv:hep-ph/0407103.

\end{thebibliography}

\end{document}